\documentclass{article}

\usepackage{preamble}

\title{Discovering heavy neutrino-antineutrino oscillations at the $Z$-pole}

\author[basel]{Stefan Antusch \email{stefan.antusch@unibas.ch}}
\author[lisboa]{Jan Hajer \email{jan.hajer@tecnico.ulisboa.pt}}
\author[lisboa]{Bruno M.\ S.\ Oliveira \email{b.m.silva.oliveira@tecnico.ulisboa.pt}}

\affiliation[basel]{Departement Physik, Universität Basel, Klingelbergstrasse 82, CH-4056 Basel, Switzerland}
\affiliation[lisboa]{Centro de Física Teórica de Partículas (CFTP), Instituto Superior Técnico (IST), Universidade de Lisboa, 1049-001 Lisboa, Portugal}

\AtBeginDocument{
  \hypersetup{
    pdftitle    = {\thetitle},
    pdfsubject  = {High Energy Physics},
    pdfauthor   = {Stefan Antusch, Jan Hajer, and Bruno M. S. Oliveira},
    pdfkeywords = {Baryon/Lepton Number Violation, Specific BSM Phenomenology},
  }
}

\begin{document}

\maketitle

\begin{abstract}
  Collider-testable type~I seesaw extensions of the \SMlong are generally protected by an approximate \LN symmetry.
  Consequently, they predict pseudo-Dirac \HNLs composed of two nearly degenerate Majorana fields.
  The interference between the two mass eigenstates can induce \NNOs leading to observable \LNV, even though the \LN symmetry is approximately conserved.
  These \NNOs could be resolved in long-lived \HNL searches at collider experiments, such as the proposed \FCCee or \CEPC.
  However, during their $Z$-pole runs, the \LN carried away by the light (anti)neutrinos produced alongside the \HNLs prevents \LNV from being observed directly.
  Nevertheless, \NNOs materialise as oscillating signatures in final state distributions.
  We discuss and compare a selection of such oscillating observables, and perform a \MClong simulation to assess the parameter space in which \NNOs could be resolved.
\end{abstract}

\clearpage

\tableofcontents

\listoffigures

\clearpage

\section{Introduction}

The detection of light neutrino flavour oscillations \cite{Super-Kamiokande:1998kpq,Super-Kamiokande:1998uiq,SNO:2001kpb,SNO:2002tuh,KamLAND:2002uet,K2K:2004iot,MINOS:2006foh,T2K:2015sqm} is a clear evidence of physics beyond the \SM.
Type~I seesaw models provide an elegant extension to the \SM, capable of explaining all observations in the neutrino sector \cite{Minkowski:1977sc,Yanagida:1979as,Gell-Mann:1979vob,Mohapatra:1979ia,Schechter:1980gr,Schechter:1981cv}.
These models introduce at least two heavy Majorana neutrinos \cite{Majorana:1937vz} whose mass terms and interactions violate the \LNLS resulting from the generalisation of the \LN observed in the \SM.
The symmetry would be recovered for two mass-degenerate \HNLs with Yukawa interactions that are identical in size and differ by a factor of the imaginary unit.
The pair of Majorana \HNLs would then form a pure Dirac \HNL, leaving the light neutrinos massless.

Current experimental constraints on the light neutrino masses~\cite{Planck:2018vyg,eBOSS:2020yzd,KamLAND-Zen:2022tow,KATRIN:2019yun,KATRIN:2021fgc,KATRIN:2021uub} impose strong restrictions on the type~I seesaw parameter space.
The additional requirements that \HNLs are detectable at collider experiments and that the model does not rely on large accidental cancellations, limits all but \SPSSs, see \cite{Antusch:2022ceb} for a more detailed discussion.
In such low-scale models, the small \LNLS breaking terms simultaneously generate light neutrino masses and a mass splitting between the two Majorana \DOFs, thereby transforming the Dirac \HNL into a so-called pseudo-Dirac \HNL.
We will focus on this scenario in this paper, but also discuss the limits in which the pseudo-Dirac \HNL approaches a pure Dirac \HNL or two separate and non-interfering Majorana \HNLs.

Similarly to nearly degenerate neutral mesons, the mass splitting in pseudo-Dirac \HNLs can induce \NNOs, see \cite{Antusch:2020pnn} and references therein, as well as the more recent discussions in \cite{Antusch:2022ceb,Antusch:2023nqd}.
The resulting oscillations between events that expose \LNC and \LNV as a function of the \HNL lifetime have previously been shown to be detectable in di-lepton channels at the \LHC \cite{Antusch:2022hhh}, see also \cite{Antusch:2017ebe}.
Moreover, a recent study \cite{Antusch:2023jsa} targeting the \FCCee showed how \NNOs could manifest as an oscillatory signature in final state distributions like the \FBA.
The present work extends these results by considering alternative observables to the oscillating \FBA and by quantifying the capability of lepton colliders such as the \FCCee \cite{FCC:2018evy} and the \CEPC \cite{CEPCStudyGroup:2023quu} to resolve \NNOs during their $Z$-pole run.

The paper is organized as follows:
\Cref{sec:low-scale seesaw} reviews the \SPSS, which captures the most relevant features of pseudo-Dirac \HNLs in low-scale type~I seesaw models, and motivates the existence of \NNOs.
\Cref{sec:final state distributions} discusses how \NNOs give rise to oscillatory signatures in final state distributions that are sensitive to the difference between processes that expose \LNCV.
\Cref{sec:simulation analysis,sec:statistics} introduce a \MC simulation and statistical analysis capable of assessing the presence of oscillations at the \FCCee and \CEPC.
\Cref{sec:results} presents the results of a scan over the three dimensional parameter space and discuss the prospects for the study of \NNOs at the \FCCee's \IDEA.
The conclusions are presented in \cref{sec:conclusion}.

\section{\sentence\NNOslong in low-scale seesaws} \label{sec:low-scale seesaw}

\resetacronym{NNO}

Low-scale type~I seesaw models generically predict pairs of nearly mass degenerate Majorana \HNLs with similar Yukawa couplings, such that \LN is an approximate symmetry of the theory.
The nature of \HNLs is then referred to as pseudo-Dirac or, synonymously, as quasi-Dirac.
This marks a paradigm shift from high scale type~I seesaw models, which explain the lightness of the known neutrino masses through the large mass of the \HNLs, as the smallness of neutrino masses is ensured by the approximate \LNLS.

In the following we briefly review a minimal benchmark scenario, the \SPSS and its phenomenological implementation, referred to as the \pSPSS, which defines the parameters necessary to capture the main features of pseudo-Dirac \HNLs relevant for collider phenomenology.
We then discuss the phenomenon of \NNOs, characteristic for pseudo-Dirac \HNLs, and how it can introduce observable \LNV in collider processes.
Finally, we comment on the pure Dirac and double-Majorana limits of pseudo-Dirac \HNLs.

\subsection{The \SPSSlong}

\resetacronym{SPSS}
\resetacronym{pSPSS}

In order to reproduce the observed light neutrino oscillation data, at least two additional Majorana fermions $N_i$ are necessary.
The \SPSS is a \BM that considers that the two \HNLs that are most likely to be detected dominate the collider phenomenology.
Compared to the \SM Lagrangian the following additional terms describe the dynamics of these fermions
\begin{equation} \label{eq:LSPSS}
\mathcal L_{\SPSS} = - y_{i\alpha} \widebar N_i^c \widetilde H^\dagger \ell_\alpha - m_M^{} \widebar N_1^c N_2^{} - \mu_M^\prime \widebar N_1^c N_1^{} - \mu_M^{} \widebar N_2^c N_2^{} + \hc + \dots ,
\end{equation}
where $N_i$ are left-chiral four-component spinor fields and $y_{i\alpha}$ are the Yukawa couplings between the heavy neutrinos and the \SM Higgs $H$ and lepton $\ell_\alpha$ doublets.
The dots summarise possible additional contributions to the light neutrino masses, \eg from additional \HNLs, which are assumed to be negligible for collider studies.
After \EWSB, the mass matrix of the neutral fermions in the flavour basis $(\vec \nu, N_1, N_2)^\trans$ is given by
\begin{equation} \label{eq:mass-matrix}
M_n =
\begin{pmatrix}
0 & \vec m_D^\trans & \vec \mu_D^\trans \\
\vec m_D & \mu^\prime_M & m_M^{} \\
\vec \mu_D^{} & m_M^{} & \mu_M^{}
\end{pmatrix} ,
\end{equation}
where the masses $\vec m_D^{} = v \vec y_1$ and $\vec \mu_D^{} = v \vec y_2$ are generated by the \SM Higgs \VEV $v \approx \unit[174]{GeV}$.

\begin{figure}
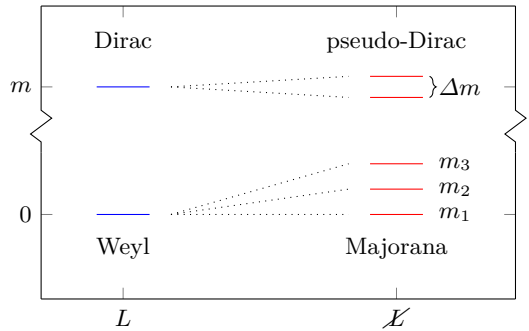

  \includepgf{splitting}
  \caption[Breaking of a \LNLSlong leads to mass splittings]{
    As long as the \LNLS stays unbroken the \SM neutrinos remain massless Weyl fermions and the heavy neutrinos form a pure Dirac \HNL.
    Light neutrino masses are only generated by the small breaking of the symmetry, which necessarily introduces a small mass splitting between the heavy neutrinos that form a pseudo-Dirac \HNL.
  } \label{fig:splitting}
\end{figure}

\resetacronym{LNLS}

When $\mu_M^{}$, $\mu_M^\prime$, and $\vec \mu_D^{}$ are simultaneously taken to zero the Lagrangian obeys an exact \LNLS and the mass matrix \eqref{eq:mass-matrix} describes the mixing of three massless neutrinos and a single heavy Dirac \HNL.
However, once a small amount of symmetry breaking is introduced by tiny mass parameters $0 < \mu_M^{}$, $\mu_M^\prime$, $\abs{\vec \mu_D^{}} \ll m_M^{}$, the light neutrinos become massive and the two heavy neutrino \DOFs $n_{\nicefrac45}$ develop a small mass splitting $\Delta m \ll m_M^{}$.
The masses of the resulting pseudo-Dirac \HNL are then
\begin{equation}
m_{\nicefrac45}^{} \approx m_M^{} \left(1 + \frac12 \abs{\vec \theta}^2\right) \pm \frac12 \Delta m ,
\end{equation}
where the active-sterile mixing parameter is defined as
\begin{equation}
\vec \theta = \flatfrac{\vec m_D}{m_M^{}} .
\end{equation}
The impact of the small symmetry breaking mass terms on the neutrino masses is depicted in \cref{fig:splitting}.

Although the precise relation between $\Delta m$ and the small mass parameters differs between different types of low-scale type~I seesaw models, \cf \cite{Antusch:2017ebe,Antusch:2022ceb}, it can be taken to be an effective parameter governing the amount of \LNV present in the model.
Remarkably, in minimal linear seesaw models, where only two \HNLs are present and the only non-zero \LN breaking term is $\mu_D$, $\Delta m$ can be expressed in terms of the light neutrino mass splittings, \ie
\begin{align} \label{eq:minimal linear seesaw}
  \Delta m_{\IO} &= m_2 - m_1 \approx \unit[743]{\mu eV} ,
  &
  \Delta m_{\NO} &= m_3 - m_2 \approx \unit[41.46]{meV}
\end{align}
for the \IO and the \NO.

More generally, the average neutrino mass $m_N$, the active-sterile mixing $\vec \theta$ and the mass splitting $\Delta m$ can be taken as free parameters.
\footnote{
  Note that due to the dots in Lagrangian \eqref{eq:LSPSS} the active-sterile mixing angle is indeed a free parameter whereas in more minimal models its components are subject to constraints from the light neutrino masses and the \PMNS parameters.
}
This philosophy is implemented by the \pSPSS \cite{Antusch:2022ceb,FR:pSPSS}.

\subsection{\sentence\NNOslong}

For an exact \LNLS the two Majorana \DOFs form a Dirac field and the heavy neutrino $N$ and antineutrino $\widebar N$ interaction eigenstates are those produced alongside antileptons and leptons and decaying into leptons and antileptons, respectively.
In this case, the particle-antiparticle nature of \HNLs is also preserved as they propagate, and their decay takes place from the same interaction state as their production, thus conserving \LN.

\begin{figure}
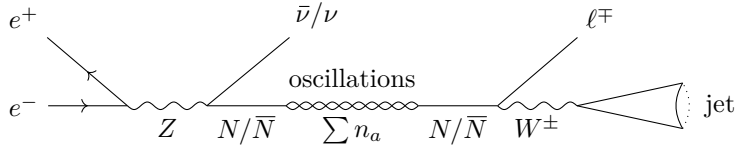

  \includepgf*{z-pole-diagram}
  \caption[Feynman diagram of the production, oscillation, and decay of a pseudo-Dirac $\HNL$]{
    Feynman diagram depicting the production, oscillation, and decay of a \HNL.
    During the $Z$-pole run at the \FCCee or \CEPC, the $e^+e^-$ collision produces an on-shell $Z$-boson, which then decays into a heavy neutrino (antineutrino) interaction eigenstate $N$ ($\widebar N$), itself a linear combination of the heavy mass eigenstates~$n_a$, alongside a light antineutrino (neutrino).
    The interference of the propagating mass eigenstates induces time dependent oscillations between $N$ and $\widebar N$.
    Due to the oscillations, the \HNL can decay into a lepton or antilepton rendering the total process $\LNCV$.
    For the parameters under consideration, the two final quarks, originating from the $W$-boson's hadronic decay, immediately hadronise into a single jet.
  } \label{fig:feynman oscillation}
\end{figure}

However, if the symmetry is broken, the mass splitting between the heavy mass eigenstates $n_a$ causes their interference as they propagate, which is reflected in oscillations between the respective particle and antiparticle interaction eigenstates.
In the formalism of \QFT with external wave packets, the oscillation probability has been derived in \cite{Antusch:2020pnn}.
As the interaction eigenstates oscillate in proper time $\tau$, so do the probabilities of their decay sourcing \LNCV processes, which are given at \LO by \cite{Antusch:2022ceb,Antusch:2020pnn}
\begin{equation} \label{eq:oscillations}
P^\text{osc}_{\lncv}(\tau) = \frac{P_+^\text{osc}(\tau) \pm P_-^\text{osc}(\tau)}{2} ,
\end{equation}
where the average probability is constant and the difference of the probabilities contains the oscillation
\begin{align} \label{eq:oscillation difference}
P_+^\text{osc}(\tau) &= 1 , &
P_-^\text{osc}(\tau) &= e^{-\lambda} \cos \Delta m \tau .
\end{align}
An additional effective damping parameter $\lambda$ is introduced to describe the loss of coherence in propagating mass eigenstates \cite{Antusch:2023nqd}.
Such a process, including the oscillations, is shown in \cref{fig:feynman oscillation}.

\begin{figure}
  \includepgf{general-shape-fcc}
  \caption[Theory predictions for the number of detected $\HNLs$ at the $Z$-pole]{
    The testable parameter space in displaced vertex searches is constrained by the geometry of the detector.
    The minimal displacement cut limits the mass range from above (\ref*{tikz:general-shape-fcc:prompt}), while the maximal displacement of the reconstructed vertices limits the mass range from below (\ref*{tikz:general-shape-fcc:outside}).
    Finally, the cross section of the model imposes a lower limit on the testable couplings (\ref*{tikz:general-shape-fcc:production}).
    The corresponding analytic approximations are given in \cite{Drewes:2022rsk}.
  } \label{fig:decay shape}
\end{figure}

Independently, heavy neutrinos decay according to an exponential decay law with its decay width $\Gamma$, which has the \PDF
\begin{equation}
P^\text{dec}(\tau) = \Gamma e^{-\Gamma\tau} .
\end{equation}
For long-lived \HNLs, the interplay between the survival probability, the detector geometry, and the dependence of the production cross section on the \HNL mass and coupling square leads to a typical shape \cite{Drewes:2022rsk}, which is presented in \cref{fig:decay shape}.

The probability of an \HNL eigenstate surviving or oscillating until it decays, thereby conserving or violating \LN, is therefore given by the product
\begin{equation}
P_{\lncv}(\tau) = P^\text{dec}(\tau) P^\text{osc}_{\lncv}(\tau) .
\end{equation}
Integrating over the time dependence yields the total probabilities for \LNCV decays after oscillations
\begin{equation}
P_{\lncv} = \int_0^\infty P_{\lncv}(\tau) \d \tau .
\end{equation}
The total ratio of events with \LNV to those with \LNC is then \cite{Antusch:2023nqd}
\begin{equation} \label{eq:R}
R := \frac{P_{\lnv}}{P_{\lnc}} = 1 - \frac{2}{1 + (1 + \nicefrac{\Delta m^2}{\Gamma^2}) \exp\lambda}
\xrightarrow{\lambda \to 0} \frac{\Delta m^2}{\Delta m^2 + 2\Gamma^2} .
\end{equation}
The limit of vanishing damping was shown to provide an accurate description of \NNO at the \LHC when the \HNLs are long-lived \cite{Antusch:2023nqd}.
Although the extension of this result to lepton colliders would require an in-depth study, it will be assumed to hold for the remainder of this work, see also the discussion in \cite{Antusch:2023jsa}.

\subsection{The pure Dirac and double-Majorana limits} \label{sec:Limits}

Neglecting damping effects, the appearance of \LNV is governed by the relation between the mass splitting and the decay width.
If the former is much smaller than the latter, the decay dominates over oscillations, and the \LNLS is approximately recovered, leading to a vanishing \LNV ratio $R \to 0$.
In this limit the processes are almost always \LNC with $P_{\lnc} \approx 1$ and $P_{\lnv} \approx 0$ and consequently the pseudo-Dirac \HNLs behave approximately like pure Dirac \HNLs.

In the other extreme, where a significant amount of oscillations take place before the heavy neutrinos have the chance to decay, the interaction states at production and decay are nearly independent.
\LNV is then maximised as $P_{\lnc} \approx P_{\lnv} \approx \frac12$ and $R \to 1$.
When the oscillations are so fast that they cannot be resolved at colliders, the pseudo-Dirac \HNLs behave approximately as two non-interfering Majorana \HNLs.
Likewise a strong damping of the oscillations due to decoherence $\lambda \gg 1$ can equally lead to $R \to 1$ \cite{Antusch:2023nqd}, as seen in definition \eqref{eq:R}.

It should be noted, however, that the mass splitting, as suggested \eg by minimal low-scale seesaw models, is typically still too small to be resolved in resonance searches at colliders.
We refer to such a scenario as being \emph{double-Majorana}.
\footnote{
Only when the mass splitting is large one can speak of two separate Majorana states.
However this situation tends to generate too heavy masses for the light neutrinos.
One can distinguish both cases by considering the ratio of the production cross section and the decay rate, since this ratio is proportional to the number of Majorana \DOFs of the measured \HNL.
}

\section{Oscillations in final state distributions} \label{sec:final state distributions}

\begin{figure}
  \begin{panels}{2}
    \includepgf*{diagram-lab-frame}
    \caption{$Z$-boson rest frame.} \label{fig:diagram-lab-frame}
    \panel
    \includepgf*{diagram-hnl-rest-frame}
    \caption{\HNL rest frame.} \label{fig:diagram-hnl-rest-frame}
  \end{panels}
  \caption[Depiction of the final state observables]{
    Depiction of the forward-backward angle \eqref{eq:FB cos} and the lepton momentum \eqref{eq:lepton momentum} in the $Z$-boson rest frame as well as the lepton opening angle \eqref{eq:cos opening angle} in the \HNL rest frame in panels \subref{fig:diagram-lab-frame} and \subref{fig:diagram-hnl-rest-frame}, respectively.
  } \label{fig:lab and hnl rest frames diagram}
\end{figure}

As we will discuss below, there are conceptionally different final state distributions sensitive to \LNV, such as \eg the \FBA caused by the polarisation of the $Z$-boson, and the \OAA caused by the polarisation of the \HNL.
While the former leads to an asymmetry in the direction in which the heavy (anti)neutrinos are emitted in the laboratory frame, the latter leads to an asymmetry of their decay products in the \HNL rest frame.
Furthermore, the polarisation of the \HNL also manifests itself in other, potentially more easily accessible observables, such as the momentum-spectrum of the charged lepton in the laboratory frame, which also shows a distinct difference between \LNCV processes.
These observables in their respective frames of reference are illustrated in \cref{fig:lab and hnl rest frames diagram}.

\subsection{\sentence\FBAlong in the decay of the $Z$-boson}

\begin{figure}
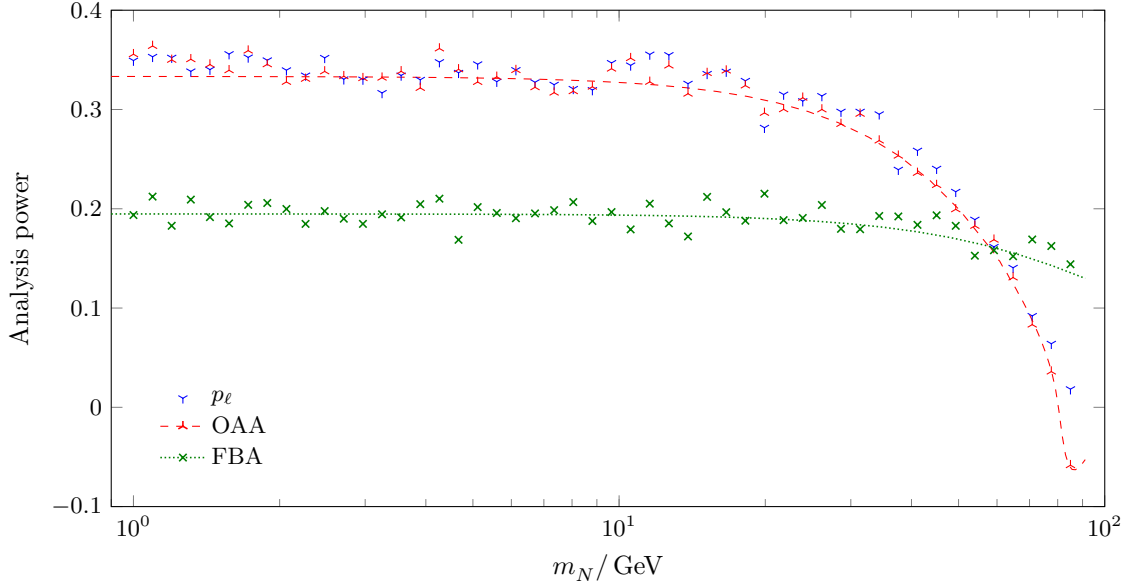

  \includepgf{apower}
  \caption[Comparison of the analysis powers]{
    Comparison between the theoretical calculation and the \MC simulation for the analysis powers of the \FBA \eqref{eq:FBA average deviation} and the \OAA \eqref{eq:OAA average and deviations}.
    The analytical results are given in \eqref{eq:analysis power FBA,eq:analysis power OAA}.
    Additionally, the \MC result for the analysis power of the lepton momentum modulus \eqref{eq:lepton momentum} calculated using the integral \eqref{eq:experimental analysis power} is shown.
  } \label{fig:apower}
\end{figure}

The $Z$-bosons produced in lepton colliders are polarised due to parity violation.
Consequently, the production cross section of an \HNL is modulated as a function of the angle between the \HNL and the momentum of the initial electron in the rest frame of the $Z$-boson and therefore the laboratory frame \cite{Lampe:1995xb,delAguila:2005pin,Blondel:2021mss}
\begin{equation} \label{eq:FB cos}
\beta := \theta_e^N(m_Z).
\end{equation}
This angle is depicted in \cref{fig:diagram-lab-frame}.
The angular-dependent \PDF for the production of a heavy neutrino~$N$ and heavy antineutrino~$\widebar N$ can be written as \cite{Blondel:2021mss}
\begin{equation} \label{eq:FBA Probability}
P_{\nicefrac{N}{\widebar N}}(\cos\beta) = P_+(\cos \beta) \pm P_-(\cos \beta).
\end{equation}
The difference between these two distributions is shown for one \HNL mass in \cref{fig:integrated-observables:fba}.
The average \PDF and the distribution of the deviations are given by
\begin{align} \label{eq:FBA average deviation}
P_+(\cos \beta) &= \frac34 \frac{(1 + \cos^2 \beta) m_Z^2 + (1 - \cos^2 \beta) m_N^2}{2 m_Z^2 + m_N^2}, &
P_-(\cos \beta) &= \frac{b}2 \cos \beta ,
\end{align}
where $b$ captures the analysis power of the \FBA asymmetry.
See \cref{fig:apower} for the dependence of the analysis power on the \HNL mass and \eqref{eq:analysis power FBA} for the analytical expression.

\begin{figure}
  \begin{panels}{3}
    \includepgf{integrated-observables-fba}
    \caption{\FBA} \label{fig:integrated-observables:fba}
    \panel
    \includepgf{integrated-observables-oaa}
    \caption{\OAA} \label{fig:integrated-observables:oaa}
    \panel
    \includepgf{integrated-observables-pell}
    \caption{Lepton momentum modulus} \label{fig:integrated-observables:pell}
  \end{panels}
  \caption[Examples of the asymmetric \PDFslong]{
    \PDFs for the \FBA \eqref{eq:FBA Probability} in panel \subref{fig:integrated-observables:fba}, the \OAA \eqref{eq:OAA Probability} in panel \subref{fig:integrated-observables:oaa}, and the modulus of the charge lepton momentum in panel \subref{fig:integrated-observables:pell}.
    The distributions are shown for the mass of the parameter point \eqref{eq:benchmark-model}.
    While the \FBA distinguishes between anti(neutrinos) the \OAA and the modulus of the charged lepton momentum distinguish between \LNCV events.
  } \label{fig:integrated-observables}
\end{figure}

Once produced, the \HNL propagates until it decays into a charged lepton or antilepton final state.
One can obtain decays into \eg $\ell^-$ either by producing an~$N$ that survives or an~$\widebar N$ that oscillates into an~$N$, thereby introducing \LNV, and analogously for $\ell^+$ final states.
This yields oscillating \PDFs for the observation of a final state $\ell^\mp$ when the \HNL decay happens after a lifetime~$\tau$ \cite{Antusch:2023jsa},
\begin{equation}
P_{\ell^\mp}(\cos\beta, \tau) = P_{\nicefrac{N}{\widebar N}}(\cos\beta) P_{\lnc}(\tau) + P_{\nicefrac{\widebar N}{N}}(\cos\beta) P_{\lnv}(\tau) ,
\end{equation}
where the oscillation probabilities are given by \eqref{eq:oscillations}.
Since the \PDF of a forward pointing lepton corresponds to the \PDF of a backward pointing antilepton
\begin{equation}
P_{\ell^-}(\cos\beta, \tau) = P_{\ell^+}(-\cos\beta, \tau) ,
\end{equation}
they can be combined to yield
\begin{equation}
P(\cos\beta, \tau) = \frac{P_{\ell^-}(\cos\beta, \tau) + P_{\ell^+}(-\cos\beta, \tau)}2 .
\end{equation}
In terms of the average \PDF and the distribution of the deviations \eqref{eq:FBA average deviation}, this \PDF reads
\begin{equation}
P(\cos\beta, \tau) = \left[P_+(\cos \beta) + P_-(\cos \beta) P_-^\text{osc}(\tau)\right] P^\text{dec}(\tau) ,
\end{equation}
where the difference between the \LNCV oscillation probabilities is defined in \eqref{eq:oscillation difference}.
From this distribution, one can derive various observables, for instance the oscillating \FBA, for details see \cite{Antusch:2023jsa}.

\paragraph{Pure Dirac and double-Majorana limits}

After integrating over the proper time the \PDF becomes
\begin{equation}
P(\cos\beta) = \int P(\cos\beta, \tau) \d \cos\beta = P_+(\cos\beta) + P_-(\cos\beta) \frac{\Gamma^2}{\Gamma^2 + \Delta m^2} e^{-\lambda} ,
\end{equation}
and the pure Dirac and the double-Majorana limits, \cf \cref{sec:Limits}, result in
\begin{align}
P^D_{\ell^\mp}(\cos\beta) &= P_{\nicefrac{N}{\widebar N}}(\cos\beta) , &
P^M_{\ell^\mp}(\cos\beta) &= P_+(\cos \beta).
\end{align}

\subsection{\sentence\OAAlong in the decay of the \HNLlong}

Another final state distribution asymmetry arises due to the polarisation of the heavy antineutrino~$\widebar N$ or heavy neutrino~$N$, which is emitted from the~$Z$ decay together with a light neutrino~$\nu$ or antineutrino~$\widebar \nu$.
Also in this case we consider the heavy neutrino production channel via the $Z$-boson, and its decay into a $W$-boson and a charged lepton, as shown in \cref{fig:feynman oscillation}.

The cross section features a characteristic dependence on the opening angle between the two leptons that are directly interacting with the \HNL during its production and decay, \ie the light neutrino and the charged lepton, in the \HNL's rest frame
\begin{equation} \label{eq:cos opening angle}
\alpha := \theta_\ell^\nu(m_N) .
\end{equation}
As depicted in \cref{fig:diagram-hnl-rest-frame} this angle can also be measured by considering the angle between the $Z$- and $W$-boson in the same rest frame.
This dependence differs between \LNCV events.
This \OAA has so far not been discussed for the $Z$-pole run of the \FCCee.
For discussions of similar observables at the \LHC see \eg \cite{Han:2012vk, Arbelaez:2017zqq} and for higher \COM energies at lepton colliders see \eg \cite{Mekala:2022cmm}.

The integration over the process presented in \cref{fig:feynman oscillation} shows that the \PDF for the observation of a \LNCV process is given by
\begin{equation} \label{eq:OAA Probability}
P_{\lncv}(\cos \alpha) = P_+(\cos \alpha) \pm P_-(\cos \alpha) .
\end{equation}
The difference of these \PDFs is shown for one example in \cref{fig:integrated-observables:oaa}.
The average \PDF and the distribution of their deviations are
\begin{align} \label{eq:OAA average and deviations}
P_+(\cos \alpha) &= \frac12 , &
P_-(\cos \alpha) &= - \frac{a(m_N, m_W, m_Z)}2 \cos \alpha ,
\end{align}
where $a$ is the analysis power.
The dependence of the analysis power on the \HNL mass is presented in \cref{fig:apower} and the analytical expression is given in \eqref{eq:analysis power OAA}.
Note that, in contrast to the \FBA distribution, there is no difference between the distributions in final states with $\ell^+$ and $\ell^-$.
The \PDFs differ only due to the \LNCV nature of the whole process.

Combining this with the oscillation probability, which introduces \LNV into the process,
yields oscillating \PDFs for the angle of the emitted charged leptons in the \HNL rest frame, as function of the lifetime $\tau$ when the \HNL decay happens,
\begin{equation}
P(\cos \alpha, \tau) = P_{\lnc}(\cos \alpha) P_{\lnc}(\tau) + P_{\lnv}(\cos \alpha) P_{\lnv}(\tau) .
\end{equation}
In terms of the average \PDF and the distribution of the deviations \eqref{eq:OAA average and deviations}, it reads
\begin{equation}
P(\cos \alpha, \tau) = \left[P_+(\cos \alpha) + P_-(\cos \alpha) P_-^\text{osc}(\tau) \right] P^\text{dec}(\tau) .
\end{equation}
The oscillations take place between the \LNC distribution $P_{\lnc}(\cos \alpha)$ and the \LNV one $P_{\lnv}(\cos \alpha)$.

\paragraph{Pure Dirac and double-Majorana limits}

After integrating over the time dependence
\begin{equation}
P(\cos \alpha) = \int P(\cos \alpha, \tau) \d x = P_+(\cos \alpha) + P_-(\cos \alpha) \frac{\Gamma^2}{\Gamma^2 + \Delta m^2} e^{-\lambda} .
\end{equation}
and approaching the pure Dirac limit $\Delta m \ll \Gamma$ and $\lambda \ll 1$ or the double-Majorana limit $\Gamma \ll \Delta m$, \cf \cref{sec:Limits}, the \PDFs becomes
\begin{align}
P^D (\cos \alpha) &= P_{\lnc}(\cos \alpha) , &
P^M (\cos \alpha) &= P_+(\cos \alpha) .
\end{align}

\subsection{Modulus of the final state lepton momentum}

The polarisation of the \HNL, together with \NNOs, also gives rise to oscillations of various other observables.
An interesting example is the oscillation of the modulus of the three-momentum of the final state charged lepton in the laboratory frame \ie the rest frame of the $Z$-boson
\begin{equation} \label{eq:lepton momentum}
p_\ell := \abs{\vec p_\ell(m_Z)}.
\end{equation}
The oscillations of $\abs{\vec p_\ell}$ in the laboratory frame are sourced by the oscillating \OAA and should be easier to access experimentally.

The starting point is the \OAA, \ie the distribution of the opening angle $\alpha$ of the charged lepton in the \HNL rest frame, \cf \cref{fig:diagram-hnl-rest-frame}.
Note that the light neutrino direction in the \HNL rest frame, which of course cannot be measured directly, is the same as the direction of the incoming $Z$-boson, which is opposite to the direction of the primary interaction vertex.
The Lorentz boost to the laboratory frame, \ie the $Z$ rest frame, is performed in exactly this direction.

Now consider a final state charged lepton produced with a certain three-momentum in the \HNL rest frame.
According to the \OAA, depending on the process being \LNC or \LNV, there is a preference of the \HNL three-momentum pointing in or opposite to the direction in which the Lorentz boost to the laboratory frame is applied.
This leads to a difference between the average lepton momentum in the laboratory frame for \LNC and \LNV events.

Finally, adding the \NNOs, we obtain a \PDF of the final state charged lepton's three-momentum in the laboratory frame.
Also this \PDF can be composed into a mean \PDF and the difference distribution of the \PDFs
\begin{equation} \label{eq:momentum distribution}
P_{\lncv}(p_\ell) = P_+(p_\ell) \pm P_-(p_\ell) .
\end{equation}
An example distribution is shown in \cref{fig:integrated-observables:pell} and we obtain the analysis power for this observable from the \MC data using
\footnote{
Note that due to the absolute value in this equation, statistical fluctuations do not cancel and can lead to an overestimation of the analysis power.
This can be counteracted by maximising the number of events and minimising the number of bins.
}
\begin{equation} \label{eq:experimental analysis power}
a^\prime = \sgn\operatorname{ext}\left({\textstyle \int_{p_\ell^{\min}}^{p_\ell^\prime} P_-(p_\ell) \d p_\ell}\right) \int \abs*{P_-(p_\ell)} \d p_\ell ,
\end{equation}
where the sign is taken as the one of the extremum of the cumulative distribution.
The numerical comparison with the analysis power of the \OAA \eqref{eq:OAA average and deviations} shown in \cref{fig:apower} allows to conclude that these two analysis powers agree with each other $a^\prime \approx a$.

\bigskip\noindent
For \HNL masses where the \FCCee and the \CEPC have the best sensitivities, the oscillation of the charged lepton momentum has a significantly better analysis power and therefore sensitivity than the oscillating \FBA.
Since we consider the charged lepton momentum to be easier accessible than the \OAA we focus on this observable when we analyse the sensitivity of the \FCCee to resolve the \NNOs.

\section{\sentence\MClong simulation} \label{sec:simulation analysis}

For the \MC simulation of \NNOs in angular depended variables we consider the \IDEA detector concept at the \FCCee.
The analysis follows the \MC analysis recently introduced in the context of the \LHC \cite{Antusch:2022ceb, Antusch:2022hhh} and the study of \NNOs in a \FBA at the \FCCee \cite{Antusch:2023jsa}.

\subsection{Event generation}

The \pSPSS has been implemented in \software{FeynRules} in order to capture the dominant collider effects of low-scale type~I seesaw models using the minimal necessary set of parameters \cite{Antusch:2022ceb,FR:pSPSS}.
It extends the \SM with two sterile neutrinos, which are described by five additional parameters, their Majorana mass (\code{Mmaj}) and mass splitting (\code{deltaM}), the light-heavy mixing angles (\code{theta1}, \code{theta2}, \code{theta3}), and the additional damping parameter (\code{damping}) defined in \eqref{eq:oscillations}.
The model file can then be parsed by \software[2.3]{FeynRules} \cite{Alloul:2013bka} to generate an \software{UFO} \cite{Degrande:2011ua} output, which is then imported by \software[3.5.3]{MadGraph5\_aMC@NLO} \cite{Alwall:2014hca} to simulate the parton level events.

A complete phenomenological study should account for the evolution of the luminosity and \COM energy throughout the runtime of the \FCCee.
However, since these quantities are still subject to change, it falls beyond the scope of the present work, which assumes a fixed \COM energy of $\sqrt s = \unit[90]{GeV}$ throughout the entire $Z$-pole run.

To account for the cross section's dependence on the \COM energy, an effective integrated luminosity $L_Z$ is defined such that the target number of bosons produced obeys \cite{PhysicsPerformanceMeeting2023}
\begin{equation}
  N_Z = L_Z \times \sigma_{e^+e^- \to Z} = 6\times10^{12}.
\end{equation}
Since the \COM energy under consideration lies below the $Z$-boson's mass, the cross sections for $e^+e^- \to Z$ cannot be directly computed.
Instead, the scattering $e^+e^- \to Z \to e^+e^-$ are simulated in \software{MadGraph} as
\begin{verbatim}
generate e- e+ > Z > e- e+
\end{verbatim}
The resulting cross section, together with the appropriate decay fraction \cite{ParticleDataGroup:2022pth}, is then used to estimate the cross section to be
\begin{equation}
  \sigma_{e^+e^- \to Z} = \frac{\Gamma_Z}{\Gamma_{Z \to e^+e^-}} \sigma^{}_{e^+e^- \to Z \to e^+e^-} = \frac{\unit[991.4]{pb}}{\unit[3.363]{\%}} = \unit[29.48]{nb} .
\end{equation}
This yields the effective integrated luminosity
\begin{equation} \label{eq:integrated luminosity}
  L_Z = \frac{N_Z}{\sigma_{e^+e^- \to Z}} = \frac{\num{6E12}}{\qty{29.48}{nb}} = \qty{203.5}{ab^{-1}} .
\end{equation}

Although the \SPSS does not constrain how \HNLs couple to different lepton flavours, muon channels at collider experiments provide cleaner signals than their electron and tau counterparts.
Therefore, the present work considers heavy neutrinos that couple exclusively to muons, such that the process depicted in \cref{fig:feynman oscillation} was generated using
\begin{verbatim}
generate e- e+ > Z > vmu nn, (nn > mu j j)
\end{verbatim}
where the multi-particles are defined using
\begin{verbatim}
define mu = mu+ mu-
define vmu = vm vm~
define j = g u u~ c c~ d d~ s s~ b b~
define nn = n4 n5
\end{verbatim}
This notation ensures that the \HNLs are produced on-shell without accounting for the interference effects of the two mass eigenstates.
As a result, events with \LNCV are created with equal probability, and the correct oscillating pattern is implemented before event reconstruction.
\footnote{
  Alternatively, the correct oscillating pattern could be directly incorporated in the \software{MadGraph} simulation through the patch described in \cite{Antusch:2022ceb,FR:pSPSS}.
}
Subsequently, \software[8.306]{Pythia} \cite{Bierlich:2022pfr} is called with the default configuration to hadronise and shower the parton level events.
Finally, detector effects are simulated by \software[3.5]{Delphes} \cite{deFavereau:2013fsa} with the \IDEA card.
\footnote{
  During the course of this work, a bug was identified in \software{Delphes}'s \IDEA card which caused the inclusion of high-energy muons within jets.
  This issue was addressed by importing the \texttt{EFlowFilter} from the \CMS card, which removes electrons and muons from the energy flow passed to the \texttt{PhotonIsolation}, \texttt{ElectronIsolation}, \texttt{MuonIsolation}, and \texttt{FastJetFinder} modules.
}

Computational limitations restrict the maximum number of simulated events to about $\num{125000}$ events per parameter point.
While this means fewer events than the maximally possible amount of collected data in the region of the parameter space with the largest cross section, it exceeds it for most of the domain outside current exclusion bounds.

\subsection{Signal and background selection} \label{sec:signal and background}

Prior to the reconstruction, the detector simulation in \software{Delphes} removes particles failing to meet the detector requirements.
Most importantly, it implements a muon isolation cut, which removes muons in the vicinity of jets by requiring the distance between the two to exceed a threshold of $\Delta R_{\max}=0.1$.
For this purpose, it employs the generalised-$k_t$ jet reconstruction algorithm for $e^+e^-$ collisions implemented in \software{FastJet} \cite{Cacciari:2011ma} using a jet radius of $R = 1.5$ in the anti-$k_t$ algorithm by setting $p = -1$.

The final state of the process depicted in \cref{fig:feynman oscillation} is composed of an undetectable light neutrino, a single muon and two soft quarks which immediately hadronise into a multi pion state and form a jet.
Events failing to meet this topology can therefore be excluded by requiring the final state to contain \cutlabel{cut:topology-muon}{exactly one muon} and \cutlabel{cut:topology-jet}{exactly one jet}.

\begin{figure}
  \begin{panels}{2}
    \includepgf{vertex-uncertainty-lab}
    \caption{Laboratory frame}\label{fig:vertex uncertainty lab}
    \panel
    \includepgf{vertex-uncertainty-hcm}
    \caption{\HNL rest frame}\label{fig:vertex uncertainty hnl}
  \end{panels}
  \caption[Vertex reconstruction uncertainty]{
    Dependence of the vertex reconstruction uncertainty as function of the displacement in the laboratory frame \eqref{eq:displacement} in panel \subref{fig:vertex uncertainty lab} and the lifetime of the \HNL in panel \subref{fig:vertex uncertainty hnl} on the Gaussian smearing parameter $\unit{\sigma}$.
    While the vertex reconstruction uncertainty is almost constant, the imperfect reconstruction of the \HNL's Lorentz factor produces a linearly increasing error in the proper lifetime modeled by a damping factor \eqref{eq:precision damping}.
    The latter effect becomes only relevant for vanishing vertex reconstruction errors.
  } \label{fig:vertex uncertainty}
\end{figure}

Furthermore, since the present study relies on long-lived \HNLs, their decay products are expected to emerge from a secondary vertex displaced from the primary $e^+e^-$ interaction.
As such, events without a \cutlabel{cut:displaced-muon}{displaced muon} may be discarded, with a threshold on the distance of $\unit[400]{\mu m} \leq d$ \cite{Drewes:2022rsk}, where
\begin{equation} \label{eq:displacement}
d := \abs{\vec r_\text{vertex}(m_Z)}                                                                                                                                                                                      \end{equation}
is the distance of the secondary vertex from the primary vertex in the laboratory frame.
To allow for proper event reconstruction in the presence of a displaced vertex, a muon is only considered if its origin falls within a cylinder with half the dimensions of the tracker in each direction, corresponding to \unit[1]{m} along the longitudinal and the transversal axis for the \IDEA detector.

Likewise, only events where the muon and the jet emerge from a \cutlabel{cut:displaced-jet}{single vertex} are valid.
The two are considered compatible if the latter contains at least two tracks within \unit[100]{\mu m} of the former's origin \cite{Antusch:2022hhh}, corresponding to the tracker's spatial resolution.
\footnote{
  A bug in \software{Delphes}' \IDEA card prevents jet tracks from being reliably recovered.
  This was addressed by considering its constituent particles directly.
}
In order to address the uncertainty in vertex reconstruction, the interaction vertex is randomly shifted based on a Gaussian distribution \eqref{eq:Gauss PDF} centred at the origin with a standard deviation of $\unit\sigma=\unit[300]{\mu m}$, resulting in a vertex uncertainty of
\begin{equation} \label{eq:vertex uncertainty}
\sigma(d) \approx \unit[240]{\mu m},
\end{equation}
which corresponds to the value derived in \cite{Aleksan:2024hyq}.
See \cref{fig:vertex uncertainty} for a comparison of the vertex uncertainty and the uncertainty in the proper time for different standard deviations.
Note that, while this uncertainty is constant in the displacement, an additional washout due to the imperfect reconstruction of the \HNL's Lorentz factor grows linearly with the displacement.
This washout is modelled by \eqref{eq:precision damping} and always subdominant to the vertex uncertainty.

\begin{table}
  \begin{tabular}{rrr}\toprule
                              & \multicolumn{2}{c}{Events}     \\ \cmidrule{2-3}
    Simulated                 & $497068$ & $\unit[100]{\%}$    \\ \midrule
    \ref{cut:topology-muon}   & $-56228$ & $\unit[-11.3]{\%}$  \\
    \ref{cut:topology-jet}    & $-2271$  & $\unit[-0.515]{\%}$ \\
    \ref{cut:n-mass-window}   & $-70177$ & $\unit[-16.0]{\%}$  \\
    \ref{cut:displaced-muon}  & $-5444$  & $\unit[-1.48]{\%}$  \\
    \ref{cut:displaced-jet}   & $-13812$ & $\unit[-3.81]{\%}$  \\
    \ref{cut:vertex-dir}      & $-3152$  & $\unit[-0.903]{\%}$ \\ \midrule
    Remaining                 & $345984$ & $\unit[69.6]{\%}$   \\ \bottomrule
  \end{tabular}
  \caption[Cut flow for simulated $Z$-pole signal events]{
    Cut flow for the simulated signal events of the \BM defined in \eqref{eq:benchmark-model}.
    The cuts are defined in \cref{sec:signal and background}.
  } \label{tab:cut flow}
\end{table}

The above cuts define the strategy for reconstructing the displaced vertex.
However, surviving events may still be subject to background sources, particularly $Z \to \tau^+\tau^-$ decays and atmospheric neutrinos.
Although current consensus does not expect these to be significant \cite{Antusch:2016ejd, Antusch:2016vyf, BayNielsen:2017yws}, they can be further mitigated with additional cuts.
Since the secondary \cutlabel{cut:vertex-dir}{vertex direction} should be collinear with the \HNL momentum, removing events where the distance between these two in $(\eta, \phi)$-space is $\Delta R \geq 0.1$ reduces both sources of background.
Moreover, removing events where the \HNL's mass, as reconstructed from its decay products, falls outside a $\pm \unit[2]{GeV}$ \cutlabel{cut:n-mass-window}{{}$N$ mass window} reduces the chance of misidentifying $\tau^+\tau^-$ decays.

The events passing the specified selection criteria can, therefore, be assumed to be devoid of background contamination.
\Cref{tab:cut flow} contains the sequence of cuts applied to a simulated sample of signal events of the \BM with parameters
\begin{align} \label{eq:benchmark-model}
m^{}_M &= \unit[14]{GeV}, &
\left(\theta_e^2, \theta_\mu^2, \theta_\tau^2\right) &= \left(0, 3, 0\right) \times 10^{-4}, &
\Gamma &= \unit[22.62]{\mu eV} .
\end{align}
Although the predominant causes of event loss in this region of parameter space are the absence of muons in the final state and poor reconstruction of the HNL’s mass, the other cuts can be more relevant in other parts of the \pSPSS parameter space.

\section{Statistical analysis} \label{sec:statistics}

In order to quantify the significance of oscillations within a dataset a statistical analysis is necessary.
For searches of \NNOs at the \LHC such an analysis has been developed exploiting that the \LN can be inferred by measuring the lepton charges \cite{Antusch:2022hhh}.
However, this technique is not applicable for the process discussed here.
Therefore, we develop an extension of the analysis to resolve \NNOs in final state distributions, such as those examined in \cref{sec:final state distributions}.
Additionally, we work with unbinned data instead of binned data in order to optimise the achievable significance.

For a statistical treatment it is necessary to define an alternative and a null hypothesis with and without oscillations, respectively.
Their fit to the simulated data can then be compared using a \LR test, which can subsequently be translated into a measure of the significance.

\subsection{Hypotheses}

We study the mean \PDF $P_+(x)$ and difference distribution $P_-(x)$ of the \PDFs appearing for the \FBA \eqref{eq:FBA average deviation}, the \OAA \eqref{eq:OAA average and deviations}, and the outgoing lepton's momentum \eqref{eq:momentum distribution} jointly by introducing the generic variable
\begin{equation}
  x \in \{\cos \alpha, \cos \beta, p_\ell\} .
\end{equation}
Note that the distribution $P_-(x)$ can be negative and is not a \PDF.
Nonetheless, we can define the \EV of the observable $x$ under these distributions
\begin{equation} \label{eq:anti symmetrised EVs}
  \ev{x}_\pm = \int x P_\pm(x) \d x .
\end{equation}
The null and alternative hypotheses can then be expressed in terms of these quantities.

\paragraph{Null hypothesis}

\begin{figure}
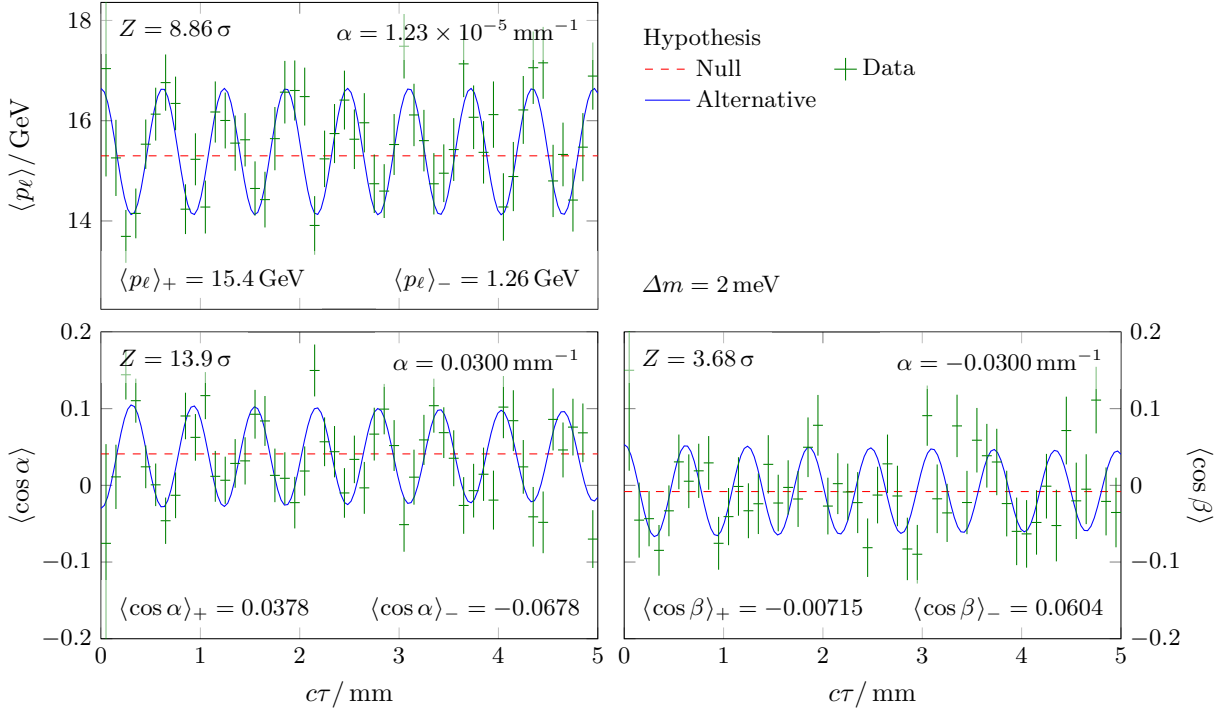

  \includepgf{fit-example}
  \caption[Example reconstruction of the oscillations]{
    Comparison of the simulated data to the null and alternative hypotheses given in \eqref{eq:hypothesis-null,eq:hypothesis-alt}, respectively.
    The data was generated using the \BM \eqref{eq:benchmark-model} with a mass splitting of $\Delta m = \unit[2]{meV}$.
    Despite the fact that the statistical analysis used here has been performed with unbinned data, for the sake of visualisation the plots show the data binned for $c\tau < \unit[5]{mm}$.
    While all the observables are best described by the alternative hypothesis, the \FBA would not allow to claim discovery since $Z\approx\unit[3]{\sigma}$ while both the \OAA and the outgoing lepton's momentum would allow for discovery at $Z>\unit[5]{\sigma}$.
  } \label{fig:fit-example}
\end{figure}

In the absence of \NNOs, \LNCV decays are equally likely independent of the \HNL's lifetime.
Consequently, the time dependent \PDF of the null hypothesis corresponds to the time independent average \PDF
\begin{equation}\label{eq:hypothesis-null}
  P_0(\tau, x) = P_+(x) .
\end{equation}
In the limit of large data samples in which statistical fluctuations vanish the null hypothesis is given by the corresponding \EV \eqref{eq:anti symmetrised EVs}
\begin{equation}
  x_0(\tau) = \ev x_+ .
\end{equation}
This hypothesis vanishes in this limit for the \FBA and the \OAA while it is finite for the lepton momentum $p_\ell$.
For realistic sample, sizes small but non-zero values are present also for the observables that are centred around zero, as show in \cref{fig:fit-example}.

\paragraph{Alternative hypothesis}

In the presence of \NNOs, the \PDF \eqref{eq:oscillations} describes a time-dependent preference for \LNCV decays.
This dependence is inherited by the \PDF of final state observables and the \PDF of the alternative hypothesis reads
\begin{equation}\label{eq:hypothesis-alt}
  P_1(\tau, x)
  = P_+(x) + P_-(x) P_-^\text{osc}(\tau) .
\end{equation}
In order to calculate the \HNL's lifetime we reconstruct its velocity and Lorentz factor.
Under the assumption that this error is independent of the displacement, the error in the reconstructed lifetime will increase for larger lifetimes.
This loss of precision can be modeled by introducing an exponential damping factor \cite{Antusch:2022hhh}
\begin{equation} \label{eq:precision damping}
  P_-^\text{osc}(\tau) \to P_-^\text{osc}(\tau) e^{-\alpha\tau} .
\end{equation}
Since the \HNL's Lorentz factor in the process depicted in \cref{fig:feynman oscillation} is fixed by kinematics, this error is subdominant and the damping factor $\alpha$ tends to be small.
In \cref{fig:vertex uncertainty} only the sample that has no additional vertex reconstruction uncertainty shows the typical behaviour of this Lorentz reconstruction error.

Finally the alternative hypothesis is given by an observable that oscillates with a frequency $\Delta m$ and an amplitude $\ev x_-$ around the average $\ev x_+$
\begin{equation}
  x_1(\tau) = \ev x_+ + \ev x_- P_-^\text{osc}(\tau).
\end{equation}
See \cref{fig:fit-example} for a comparison between the three observables discussed here using the \BM point \eqref{eq:benchmark-model}.

\subsection{\sentence\LRlong}

The preference of the data for one of the two hypotheses can be quantified using an unbinned weighted least squares fit.
The likelihood that an event with the observable $x_e$ is described by the hypotheses predicting that it measures $x_{\nicefrac10}$ at the observed proper time $\tau_e$ is given by
\begin{equation}
  L_{\nicefrac10}^e = f_\text{Gauss}(x_e, \text \sigma_e; x_{\nicefrac10}(\tau_e)),
\end{equation}
where the \PDF of the Gaussian distribution and its standardised form are
\begin{align} \label{eq:Gauss PDF}
  f_\text{Gauss}(\mu, \text \sigma; x) &= \frac{1}{\text \sigma} f_\text{Gauss}\left(\frac{x - \mu}{\text \sigma}\right), &
  f_\text{Gauss}(x) &= \frac{1}{\sqrt{2\pi}} \exp\left(-\frac{x^2}2\right) .
\end{align}
For each event the standard deviation $\sigma_e$ is calculated from the $30$ nearest neighbours within the complete dataset measured using the \HNLs' proper lifetimes.
The likelihood that the data is in agreement with the hypothesis is then the product over all events
\begin{equation}
  L_{\nicefrac10} = \prod_\text{events} L_{\nicefrac10}^e .
\end{equation}

\resetacronym{LR}

Since the null hypothesis is recovered from oscillations when $\Delta m = 0$ or $\alpha \to \infty$, the former is nested within the latter.
Assuming furthermore that both fits converge on their respective global minima, the likelihoods obey
\begin{equation}
L_0 \leq L_1 .
\end{equation}
The agreement of the data for one of the hypotheses can then be quantified by the \LR
\begin{equation}
\Lambda := \frac{L_0}{L_1} \in \mathopen]0, 1] .
\end{equation}
The lower and upper limits indicate that oscillatory behaviour is either strongly present or entirely absent in the data.

\subsection{Significance}

While the \LR provides a quantification for a dataset's agreement for one hypothesis over another, it does not account for the impact of statistical fluctuations.
As such, it is unable to provide a quantification of the significance of the alternative hypothesis by itself.
Such a statement is, instead, related to the probability $p$ of observing the same \LR under the null hypothesis.

\begin{figure}
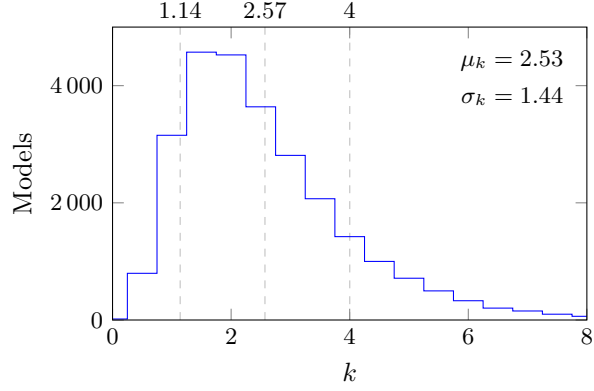

  \includepgf{dof-histogram}
  \caption[\sentence\DOFslong of the $\chi^2$-distribution]{
    Distribution of the \DOFs $k$ of the $\chi^2$ distribution with \PDF \eqref{eq:chi^2 PDF}.
    Each \DOF is the result of a fit to five $\LLR$ calculated from datasets generated according to the \PDF for one of the simulated parameter points.
    The distribution is centred at $\mu_k=2.57$ and has a standard deviation $\text{\sigma}_k=1.43$.
} \label{fig:DOFs}
\end{figure}

Larger values of the \LLR
\begin{equation}
\lambda := - 2 \ln \Lambda = \ln L^2_1 - \ln L^2_0 \in \mathbb R_0^+ ,
\end{equation}
indicate an increased preference for oscillations in the data.
It is normalised such that in the limit of large samples sizes and therefore vanishing fluctuations it follows a $\chi^2$ distribution with the \PDF
\begin{align} \label{eq:chi^2 PDF}
  f_{\chi^2}(k; \lambda) &= \frac12 f_P\left(\frac\lambda2; \frac{k-2}{2}\right), &
  f_P(\mu; n) &= \frac{n^\mu}{\Gamma(\mu+1)} e^{-n} ,
\end{align}
where $k$ are the \DOFs of the distribution.
In order to estimate the \DOFs for each parameter point additional datasets without oscillations are generated from the time-integrated \PDFs.
The \DOFs are then extracted from a fit of the $\chi^2$ distribution to five \LLRs.
The resulting distribution of the \DOFs over all parameter points is given in \cref{fig:DOFs} and has the average value of $k = 2.57 \pm 1.43$.

The probability of the hypothesis with oscillations in the \MC data can then be calculated as
\begin{equation}
  p = 1 - F_{\chi^2}(k; \lambda) ,
\end{equation}
where the \CDF of the $\chi^2$ distribution is
\begin{align}
  F_{\chi^2}(k; \lambda) &= P\left(\frac{k}{2},\frac{\lambda}{2}\right) ,
  &
  P(s, x) &= \frac{\gamma(s,x)}{\Gamma(s)} ,
  &
  \gamma(s,x) &= \int_0^x t^{s-1} e^{-t} \d t ,
\end{align}
and $P(s,x)$ is the regularized $\Gamma$-function while $\gamma(s,x)$ is the lower incomplete $\Gamma$-function.

The probability of rejecting the null hypothesis although it is true can then be converted into a significance
\begin{equation}
  F_\text{Gauss}(Z) = 1 - p ,
\end{equation}
where the \CDF of the standardised Gaussian distribution with \PDF \eqref{eq:Gauss PDF} is
\begin{equation}
  F_\text{Gauss}(x) = \int_{-\infty}^x f_\text{Gauss}(t) \d t .
\end{equation}

Therefore, for a fixed number of \DOFs, a larger \LLR corresponds to a smaller $p$-value and a greater significance, and oscillations can be interpreted as being completely absent when
\begin{equation}
  \lambda \leq k
  \iff
  \unit[50]{\%} \leq p
  \iff
  Z \leq \unit[0]{\sigma} .
\end{equation}

In the case of the examples shown in \cref{fig:fit-example} the observables are then shown to allow for the discovery of oscillations, with a significance exceeding the threshold of \unit[5]{\sigma} for the \OAA and the muon momentum.
However, the \FBA would not suffice to claim discovery.
Generally, the significance achievable with the \HNL \FBA \eqref{eq:FBA average deviation} is below that of the opening angle \eqref{eq:OAA average and deviations} and the outgoing muon's momentum \eqref{eq:momentum distribution}.
The latter two observables give nearly equal results, as they were shown to be closely related in \cref{fig:integrated-observables}.

\section{Results} \label{sec:results}

\begin{figure}
  \includepgf{significance-fixed-mmaj-theta}
  \caption[Significance as function of mass splitting, number of events, and vertex uncertainty]{
    Scan over the expected significance of the reconstructed \NNOs as a function of $\Delta m$ and the number of produced \HNLs up to an integrated luminosity of $L_Z = \unit[203.5]{ab^{-1}}$, \cf the calculation leading to \eqref{eq:integrated luminosity}.
    The data was simulated for the \BM model \eqref{eq:benchmark-model} and the vertex Gaussian smearing described in \cref{sec:signal and background} is varied from a perfect reconstruction (upper left) to the currently expected uncertainty \cite{Aleksan:2024hyq} (lower right).
    The vertical lines indicate the mass splittings corresponding to the linear seesaw with \IO and \NO \eqref{eq:minimal linear seesaw}.
    While the minimal linear seesaw with \IO is well within the discovery reach, the \NO could only be discovered with a significantly improved vertex reconstruction uncertainty.
  } \label{fig:significance-fixed-mmaj-theta}
\end{figure}

\begin{figure}
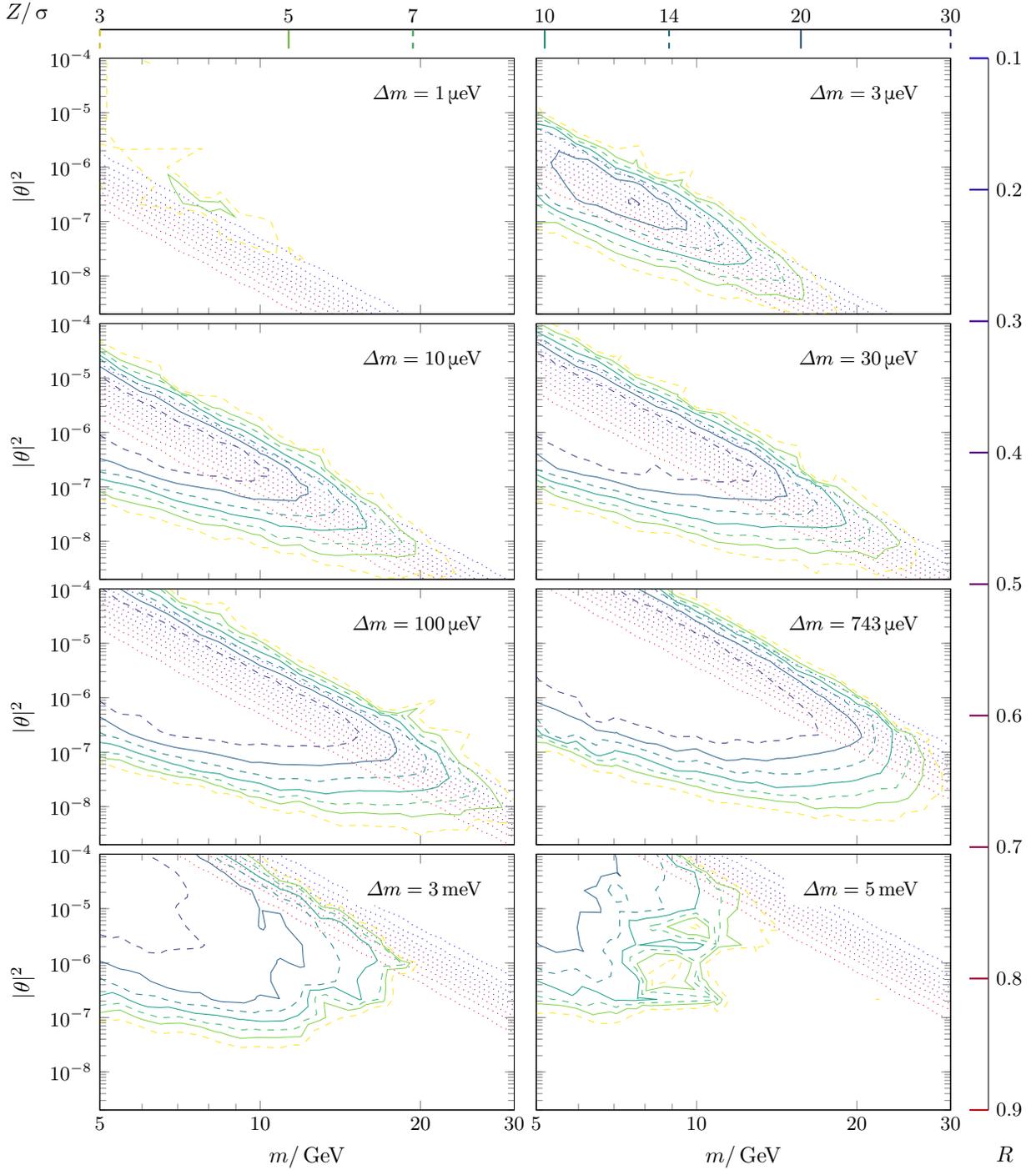

  \includepgf{significance-fixed-deltam}
  \caption[Significance as function of $\HNL$ mass, coupling, and mass splitting]{
    Scan over the expected significance of the reconstructed \NNOs as a function of \HNL mass and the active sterile mixing parameter for various mass splittings.
    The diagonal bands correspond to \LNV ratios \eqref{eq:R} of $0.1 \leq R \leq 0.9$ for the given mass splitting.
    While a mass splitting of \unit[1]{\mu eV} will be barely detectable, the \BM constituted by the minimal linear seesaw with \IO \eqref{eq:minimal linear seesaw} and leading to a mass splitting of \unit[743]{\mu eV} has an almost maximal significance reach.
    For larger mass splittings the vertex uncertainty discussed in \cref{sec:signal and background} and shown in \cref{fig:significance-fixed-mmaj-theta} becomes relevant and can be observed as a maximal reachable mass.
    For mass splitting larger than \unit[5]{meV} the significance is again drastically reduced.
  } \label{fig:significance-fixed-deltam}
\end{figure}

\begin{figure}
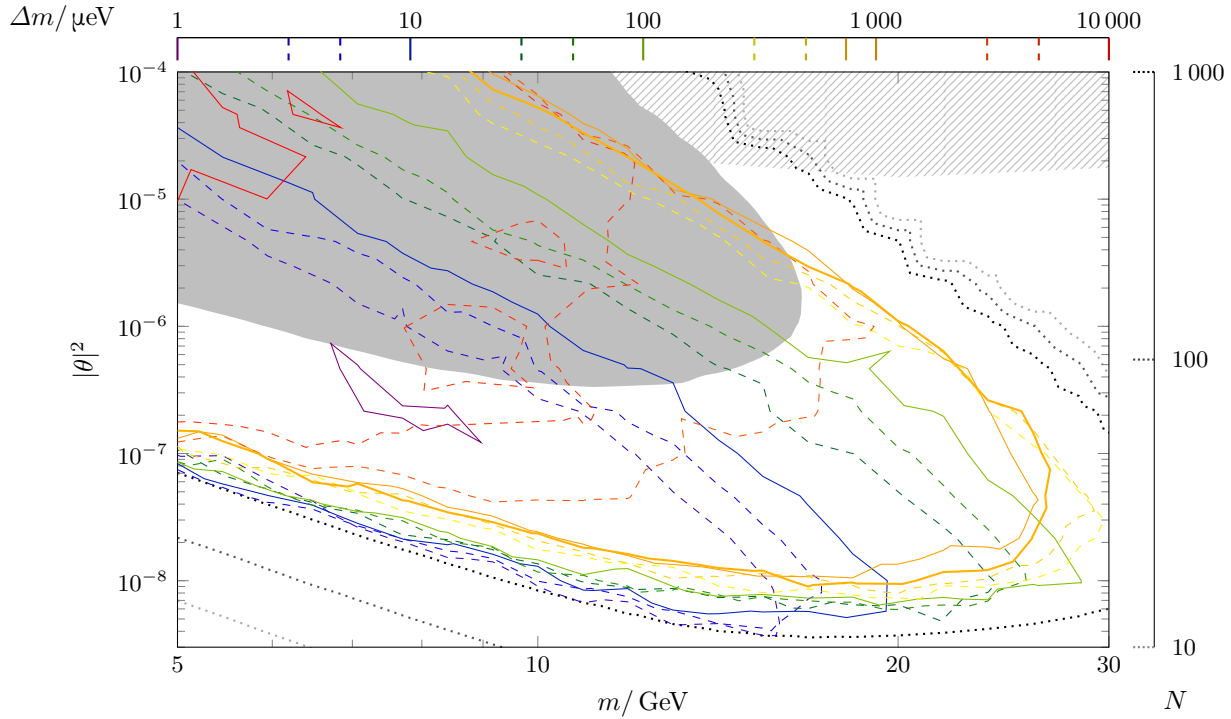

  \includepgf{significance-5sigma}
  \caption[Discovery reach as function of $\HNL$ mass, coupling, and mass splitting]{
    Comparison of the \unit[5]{\sigma} contours for different mass splitting shown as a function of the \HNL mass and the active-sterile mixing parameter.
    The thicker contour indicates the minimal linear seesaw model \eqref{eq:minimal linear seesaw} with \IO, corresponding to $\Delta m = \unit[743]{\mu eV}$.
    Exclusion boundaries from displaced vertex searches \cite{ATLAS:2020xyo,CMS:2022fut} and prompt \LNV processes \cite{ATLAS:2015gtp,CMS:2018iaf} are denoted by the grey and shaded regions, respectively.
    The dotted lines correspond to an expectation of \num{10}, \num{100}, and \num{1000} reconstructed events and demonstrate that at least around \num{1000} events need to be detected in order to claim a \unit[5]{\sigma} discovery of \NNOs.
  } \label{fig:significance-5sigma}
\end{figure}

\begin{figure}
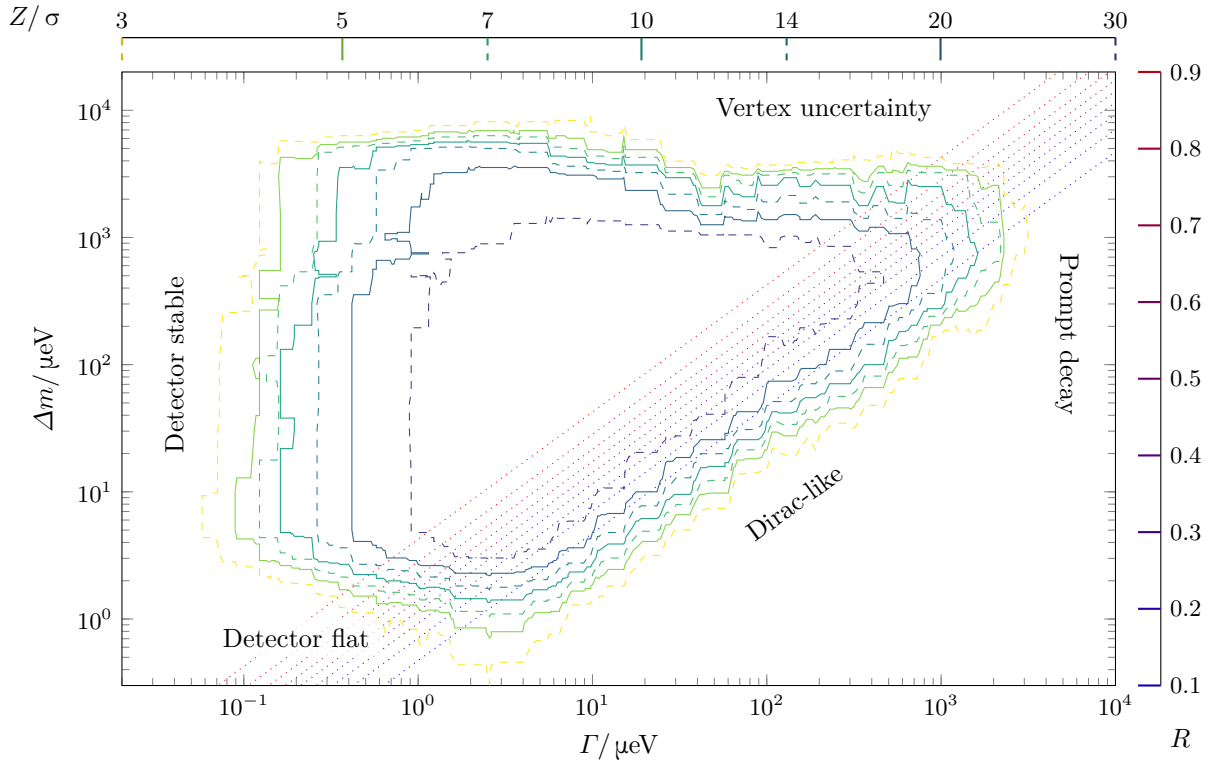

  \includepgf{significance-width-deltam}
  \caption[Significance as function of decay width and mass splitting]{
    Scan over the maximal achievable significance of the reconstructed \NNOs as a function of the decay width and the mass splitting.
    Parameter points that are excluded by past and current experiments are not included in the data and parameter points with identical decay widths and mass splittings but lower significance are not visible in this representation.
    The diagonal band corresponds to parameter points with an \LNV ratio \eqref{eq:R} of $0.1 \leq R \leq 0.9$.
    Additionally, five constraints that limit the discovery reach are indicated.
    If an \HNLs decays before the minimal distance cut it has a \emph{prompt decay}.
    Correspondingly, if an \HNL decays outside the detector geometry it is \emph{detector stable}.
    Likewise, if its oscillation length is longer than the detector geometry it becomes \emph{detector flat}.
    In the other extreme, if its oscillation length is shorter than the \emph{vertex uncertainty} its oscillation can also not be resolved.
    Finally, if its oscillation length is longer than its decay length, the oscillation is prevented from developing and the \HNL becomes \emph{Dirac-like}.
  } \label{fig:significance-width-deltam}
\end{figure}

\begin{figure}
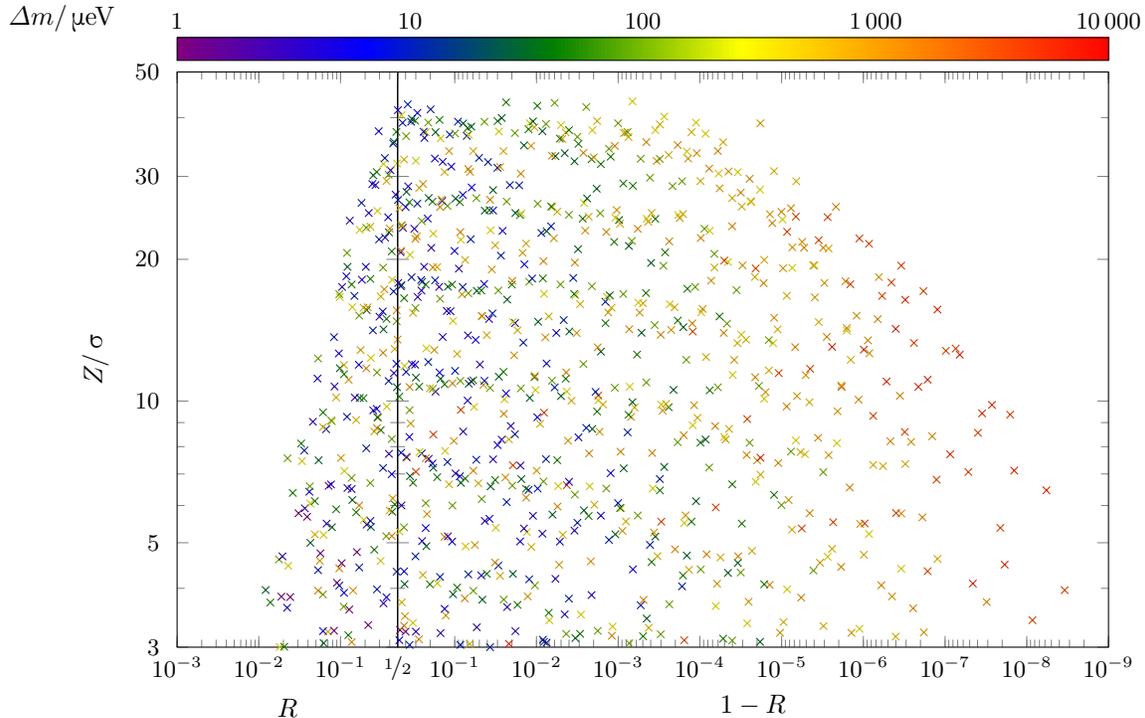

  \includepgf*{significance-rll}
  \caption[Significance as function of $\LNV$ ratio and mass splitting]{
    Achievable significance as a function of the \LNV ratio \eqref{eq:R} and the resolvable mass splitting.
    The discoverable parameter points span a range of $10^{-2} \lesssim R \lesssim 1-10^{-8}$.
  } \label{fig:significance-rll}
\end{figure}

\begin{figure}
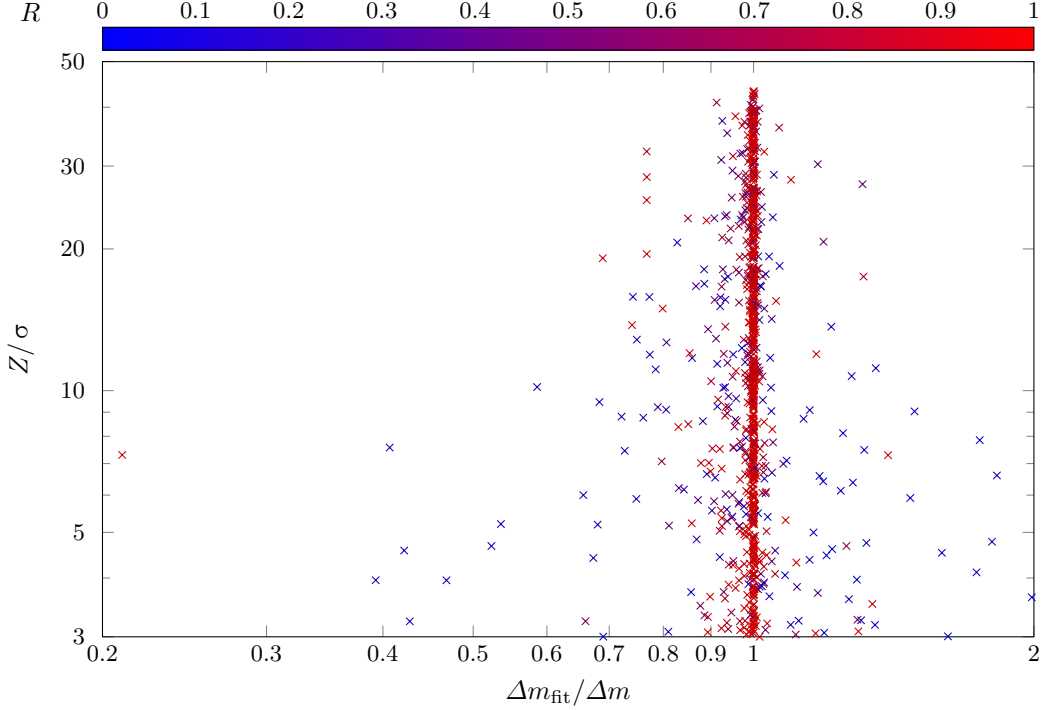

  \includepgf*{significance-error}
  \caption[Mass splitting reconstruction error as function of significance and $\LNV$ ratio]{
    Relationship between the accuracy of the reconstructed mass splitting and the achievable significance of the reconstructed \NNOs and the \LNV ratio \eqref{eq:R}.
    The plot contains \num{976} points of which \unit[88.5]{\%} are within a narrow band of $\Delta m_\text{fit}/\Delta m \in [0.9, 1.1]$.
    The majority of the points that have a larger reconstruction error are Dirac-like.
    Hence the corresponding distributions expose only enough oscillations to exclude the null hypothesis but the oscillation length is too large to reliably infer the mass splitting.
  } \label{fig:significance-error}
\end{figure}

The framework introduced in the previous sections is able to simulate and reconstruct \NNOs at the $Z$-pole of a lepton collider and assess their significance.
When employed to the \BM \eqref{eq:benchmark-model} whose oscillatory data is presented in \cref{fig:fit-example}, it showed that the \IDEA detector at the \FCCee will be able to resolve \NNOs.
This allows to explore the conditions that are necessary to discover \NNOs at such a collider experiment.
To this end, we perform a complete scan over the relevant parameters of the \pSPSS, consisting of the \HNL mass and coupling as well as its mass splitting.
Additionally, we consider the impact of varying the vertex uncertainty and the luminosity, and therefore the number of measured events as well as the beam parameters.
Finally we present the results in terms of the significance calculated as described in \cref{sec:statistics}.

\Cref{fig:significance-fixed-mmaj-theta} shows how the resolvability of \NNOs depends on the mass splitting and the number of simulated events for the \BM \eqref{eq:benchmark-model}.
Larger significances can be achieved through an increase in the number of events, with the required amount remaining approximately constant inside a given range and experiencing a steep increase beyond it.
For the \BM in question, this plateau corresponds to about \num{2000} events for a \unit[5]{\sigma} discovery.
It starts when a sufficiently large sample of \LNV events becomes available, requiring an oscillation length of the same order as the tracker geometry, $\Delta m \approx \unit[10]{\mu eV}$, and ends once the oscillation length is of the same order as the vertex uncertainty.

In the same figure, we show the impact of varying the vertex uncertainty that is in all other results fixed by \cref{eq:vertex uncertainty} and in more detail described in \cref{fig:vertex uncertainty}.
This analysis shows that, for this \BM point, the mass splitting of the minimal linear seesaw \eqref{eq:minimal linear seesaw} with \IO would be discoverable while the \NO can only be probed with a considerably reduced vertex uncertainty.

Considering the full planned runtime at the \FCCee's $Z$-pole and the current expected vertex uncertainty \cite{Aleksan:2024hyq}, \cref{fig:significance-fixed-deltam,fig:significance-5sigma} show how the resolvability of \NNOs depends on the \HNL's mass and mixing angles, for different values of the mass splitting.
For small mass splitting, the achievable significance can be interpreted as the interplay of the expected number of events, shown in \cref{fig:decay shape} and the \LNV ratio calculated in \eqref{eq:R} and shown as bands in \cref{fig:significance-fixed-deltam}.
As the oscillation lengths become comparable to the uncertainty in vertex reconstruction, this too becomes a limiting factor in the resolution of \NNOs and leads to a maximal reachable mass.
The dotted lines in \cref{fig:significance-5sigma} correspond to parameter points that lead to $10$, $100$, and \num{1000} detectable \HNL decays.
Therefore only parameter points that lead to at least around \num{1000} observable \HNL decays can lead to a discovery of \NNOs with \unit[5]{\sigma}.

\Cref{fig:significance-width-deltam} illustrates how the resolution of \NNOs depends on the mass splitting between the \HNLs and their decay width, by showing the maximally possible significance as function of these variables.
Its behaviour can be explained through the interplay of multiple constraints.
Extremely short- and long-lived \HNLs avoid detection either by failing to meet the minimum displacement threshold or by decaying outside the detector as illustrated in \cref{fig:decay shape}.
Additionally, when the oscillation length is of the order of the vertex reconstruction uncertainty \eqref{eq:vertex uncertainty}, the \NNOs become unresolvable.
Similarly, if the oscillation length becomes larger than the tracker geometry, \NNOs become undetectable.
If the decay length of the \HNLs becomes shorter than their oscillation length, oscillations can also not take place anymore and the \HNLs become Dirac-like, \cf \cref{sec:Limits}.

\Cref{fig:significance-rll} shows the achievable significance as function of the \LNV ratio \eqref{eq:R} and the mass splitting.
The relation between larger mass splittings and larger \LNV ratios is eminent.
Additionally, it is clear that larger \LNV ratios are easier to discover than smaller.
\NNOs can only be discovered if the interplay of oscillation and decay results in $10^{-2} \lesssim R \lesssim 1-10^{-8}$ and the range shows a preference for the reconstruction of higher-frequency oscillations.

Finally, \cref{fig:significance-error} permits the evaluation of the quality of the reconstruction of the mass splitting between the \HNLs.
Of the reconstructed mass splittings, \unit[88.5]{\%} lie within a narrow band of \unit[10]{\%} around the correct one.
Note that this holds although the statistical procedure presented in \cref{sec:statistics} has mainly been designed to quantify the rejection of the null hypothesis, and treats the mass splitting as a free parameter of the alternative hypothesis.

\section{Conclusion} \label{sec:conclusion}

\resetacronym{FBA}
\resetacronym{OAA}

With the light neutrinos produced together with the \HNLs being unobservable, signs of \LNV at $e^+ e^-$ colliders can only be observed in final state distributions.
We have addressed the question of which are the best observables to resolve \NNOs and extract the \HNL mass splitting during the $Z$-pole run of the \FCCee or the \CEPC.
We have shown that there are two independent angular distributions that can be exploited, the \FBA caused by the polarisation of the $Z$-boson, and the \OAA caused by the polarisation of the \HNL.
While the former leads to an asymmetry in the direction in which the heavy (anti)neutrinos are emitted, the latter leads to an asymmetry of the (anti)leptons produced from the decay of the heavy (anti)neutrinos.
We find that instead of reconstructing the \OAA angular dependence directly, a related and easier accessible observable is provided by the momentum-spectrum of the charged lepton in the laboratory frame, which shows a distinct difference between \LNCV processes, dominantly caused by the \OAA.

Using this improved observable, which turns out to be more sensitive than the previously studied \FBA, we perform a first complete exploration of the parameter space where \NNOs could be discovered at the \FCCee.
For this study we have developed an unbinned statistical analysis from first-principles.
We have shown the potential reach for the $Z$-pole run of future lepton collider such as the \FCCee and the \CEPC and have identified the main constraints that limit this reach.

\subsection*{Acknowledgment}

We thank Giacomo Polesello for helpful discussion on the vertex reconstruction at the \FCCee, the kinematics of the $Z$-pole run, and the potential of an unbinned fit over a binned fit.
The work of Jan Hajer and Bruno Oliveira was partially supported by the Portuguese \FCT through the projects CERN/\allowbreak FIS-PAR/\allowbreak0002/\allowbreak2021 and CERN/\allowbreak FIS-PAR/\allowbreak0019/\allowbreak2021, which are partially funded through POCTI (FEDER), COMPETE, QREN and the EU.

\appendix

\section{Analysis powers}

The analysis power of the \FBA appearing in \eqref{eq:FBA average deviation} is
\begin{equation} \label{eq:analysis power FBA}
b = \frac32 \frac{m_Z^2}{m_Z^2 + \flatfrac{m_N^2}{2}} \Delta \gamma,
\end{equation}
where the normalised difference between left- and right-chiral couplings of the charged leptons to the $Z$-boson is
\begin{align}
\Delta \gamma &= \gamma_L - \gamma_R, &
\gamma_{\nicefrac LR} &= \frac{g_{\nicefrac LR}^2}{g_L^2 + g_R^2}, &
g_L &= 1 - g_R, &
g_R &= 2 \sin^2 \theta_w.
\end{align}
Its dependence on the \HNL mass is depicted in \cref{fig:apower}.

The analysis power of the \OAA appearing in \eqref{eq:OAA average and deviations} is
\begin{align} \label{eq:analysis power OAA}
a &= \frac{\sigma_1}{\sigma_0} , &
\sigma^{}_{\nicefrac10} &= \sigma^p_{\nicefrac10} \sigma^d_{\nicefrac10} , &
\sigma^p_{\nicefrac10} &= r_p \mp 2,
\end{align}
where
\begin{multline}
\sigma^d_{\nicefrac10} =
\left[4 - (3 \pm 1) r_d\right] r_d \gamma
+ \left[(3 - (1 \pm 1) r_d) (1 - r_d) - \gamma^2\right] \gamma l
\\
+ \left[(2 \mp r_d) (1 - r_d)^2 -(6 - (4 \pm 1) r_d) \gamma^2\right] t
,
\end{multline}
with
\begin{align}
r_d &= \frac{m_N^2}{m_W^2}, &
r_p &= \frac{m_N^2}{m_Z^2}, &
\gamma &= \frac{\Gamma_W}{m_W},
\end{align}
and the appearing functions are
\begin{align}
l &= \ln\left(1 - r_d \frac{2 - r_d}{1 + \gamma^2}\right), &
t &= \arctan\left(r_d \gamma, 1 - r_d + \gamma^2\right).
\end{align}
Its dependence on the \HNL mass is depicted in \cref{fig:apower}.

\sloppy\printbibliography

\end{document}